\RequirePackage{fix-cm}
\documentclass[smallextended]{svjour3}       % onecolumn (second format)
\smartqed  % flush right qed marks, e.g. at end of proof
\usepackage{graphicx}
\usepackage{epstopdf}
\begin{document}
\title{Modified entropy of Kerr-de Sitter black hole in Lorentz symmetry violation theory}


\author{ Y. Onika  Laxmi        \and
  T. Ibungochouba Sing${\rm h}^{*}$ \and I. Ablu Meitei%etc.
}


\institute{ Y. Onika  Laxmi  \at
               Department of Mathematics, Manipur University, Canchipur, 795003, India \\
              \email{onikalaxmi@gmail.com}           %  \\
           \and
            T. Ibungochouba Sing${\rm h}^{*}$ \at
              Department of Mathematics, Manipur University, Canchipur, 795003, India \\
               \email{ ibungochouba@rediffmail.com}
\and 
I. Ablu Meitei \at
Physics Department, D.M. College of Science, Imphal, Manipur, 795001, India\\
\email{ablu.irom@gmail.com}}

\date{Received: date / Accepted: date}

\maketitle

\begin{abstract}
The quantum tunneling radiation of scalar particles near the event horizon of Kerr-de Sitter black hole is investigated in three systems of coordinates namely naive coordinate system, Painleve coordinate system and Eddington coordinate system using Lorentz violation theory in curved space time. The Klein-Gordon equation of scalar particles is transformed into Hamilton-Jacobi equation by using Lorentz violation theory in curved space time. We observe that due to Lorentz violation theory, the expressions of Hawking temperatures, the Bekenstein-Hawking entropies and heat capacities near the event horizon of Kerr-de Sitter black hole are modified. The Hawking temperatures, entropies and  heat capacities increase or decrease depending upon the choices of ether like vectors $u^\alpha$.
\end{abstract}

\section{Introduction}

Using quantum field theory in curved space time, Hawking [1, 2] proposed that a black hole emits radiation like a blackbody radiation. Refs. [3-5] lead to connect  black hole with thermodynamics. They showed that the entropy of black hole is proportional to its horizon area. Since then, many researchers studied the Hawking radiation  from the static, stationary and nonstationary black hole.  Refs. [6, 7] developed a new technique to investigate the thermal radiation of stationary and non stationary black holes. In this technique, the radial part derived from Dirac equation and Maxwell's equation can be recast into a single form of wave equation at the event horizon of the black hole. Then the ingoing and outgoing wave equations can be derived and the corresponding Hawking radiation of black hole can be studied. More interesting results of Hawking radiations of stationary and nonstationary black holes have been obtained in [8-11].
Taking self gravitational interaction and the change of curved space time, Refs. [12-14] proposed the modification of Hawking pure thermal radiation spectrum and they found the conservation of information during the thermal radiation from stationary black hole. {\bf Zhang and Zhao [15-17] have generalized the Parikh-Wilczek's  technique to investigate the tunneling of charged massive particle near the event horizon of black hole where the geodesic equation has been derived by studying the relation between the phase and group velocities of the tunneling particle}.
The tunneling of massless particle through quantum horizon of  Schwarzscild black hole is investigated by taking quantum gravity into account. The minimal length, minimal momentum and maximal momentum are also discussed using quantum gravity effects in the framework of the generalized uncertainty principle. There seems to be some connection between Lorentz symmetry violation and quantum gravity effects [18]. Nozari and Mehdipour [19] also investigated the Hawking radiation as a tunneling near the horizon of Schwarzscild black hole in noncommutative space time. They showed that there is no correlation between different modes of radiation in the noncommutative space time.
 Ref. [20] as an extension of complex path analysis [21],  discussed the Hawking radiation of black hole using semi classical Hamilton-Jacobi equation, Feynman prescription and WKB approximation ignoring the particle back reaction effect. The Hawking radiation of Schwarzschild black hole has been discussed in naive coordinate system and isotropic coordinate or invariant coordinate system. The Hawking radiations in different black holes have been derived using Hamilton-Jacobi equation [22-25 ]. Kerner and Mann [26] proposed the tunneling of {\rm spin-1/2} fermions near the event horizon of black hole using Dirac equation, Pauli Sigma matrices and WKB approximation. To choose appropriate Gamma matrix from the line element and  substituting to the Dirac equation, the imaginary part of the action related to the Boltzmann factor for emission at the Hawking temperature is obtained.  The tunneling of vector boson particle near the event horizon of black hole is investigated using Hamilton-Jacobi ansatz to Proca equation, Feynman presciption and WKB aproximation [27, 28]. They show that the emission temperature is consistent with the Hawking temperature corresponding to scalar particle for the Schwarzschild black hole geometry. Applying Proca equation, the Hawking  temperature of different black holes have been obtained in [29-32]. The change in Bekenstein-Hawking entropy of black hole has been investigated using Hamilton-Jacobi equation beyond the semiclassical approximation and first law of black hole thermodynamics [33]. Refs. [34-37] extended the change in Bekenstein-Hawking entropy by tunneling of fermions across the event horizon of black hole using Dirac equation in curved space time. The general theory of relativity is  based on gravity which is impossible to renormalise, therefore many researchers proposed different modified gravity theories. One of the theories which can break at high energy levels is Lorentz symmetry. Refs. [38-42] proposed the Lorentz symmetry violation based on  various gravity models. 
 Refs. [43, 44] investigated the Lorentz violation theory of Dirac equation in flat Euclidean space using ether like vectors $u^\alpha$. Many interesting results of Hawking radiation of black hole with Lorentz violation theory have also been discussed in [45-48]. {\bf Liu, et al. [49] discussed the entropy correction of black hole using the modified Hamilton-Jacobi equation in Lorentz violation theory and beyond the semiclassical approximation. They show that the modified entropy depends on the choices of ether like vector $u^\alpha$}. 
 
  A quantum theory of gravity would be required to unify the general relativity which is a classical theory and the standard model of particle Physics which is a quantum theory. The relevant scale of quantum theory of gravity is the Planck scale $(\sim 10^{19}\rm{GeV})$. It is not possible to conduct experiments at such a scale. Violation of Lorentz symmetry could provide a key to experimental probes at the Planck scale [50]. If Lorentz invariance is violated at the Planck scale, it is likely that there must be an interpolation at much lower energy so that a small amount of Lorentz violation should be present at all energies [51]. Some experiments designed to test low energy residual effects of Planck scale Lorentz violation are- (i) Penning traps [52-54] (ii) Clock comparison experiments [55, 56] (iii) Spin polarized torsion pendulum [57, 58] (iv) Meson tests [59-61] (v) Muon tests [62, 63] (vi) Photon tests [64, 65] (vii) Neutrino tests [66, 67] (viii) Gravity tests [68, 69] (ix) Cosmic rays [70, 71].

 To study the modified Hawking temperature, entropy and  heat capacity near the event horizon of KdS black hole in Lorentz violation theory, we use three coordinate systems namely naive, Painleve and Eddington coordinate systems. The Painleve coordinate system has many attractive features. Firstly, the metric components are regular near the event horizon of black hole. Secondly, the constant time slice is flat Euclidean space. The importance of the second one is that WKB approximation can be utilized to calculate the emission rate. Using the quantum tunneling of flat space time, the WKB approximation can be obtained. Thirdly, the time like Killing vector field keeps the spacetime stationary. The advantage of a new coordinate system is to eliminate the singularity near the event horizon of KdS black hole. The components of metric in new coordinate system  satisfied the Landau's condition of the coordinate clock synchronization [72-75]. 

The paper is organised as follows: In section 2, we discuss the derivation of modified Hamilton-Jacobi equation from Klein-Gordon equation using Lorentz violation theory in curved space time. In section 3, the actual Hawking temperature and the modified Hawking temperature are obtained from direct calculation and naive coordinate system using modified Hamilton-Jacobi equation with Lorentz violation theory in curved space time respectively. Using Painleve coordinates and Hamilton-Jacobi equation with Lorentz violation theory in curved space time, the Hawking temperature,  change in Bekenstein-Hawking entropy and heat capacity of KdS black hole near the event horizon is investigated in section 4. In section 5, we discuss the modified Hawking temperature, change in Bekenstein-Hawking entropy and  heat capacity of KdS black hole using Eddington coordinate and Hamilton-Jacobi equation in Lorentz violation theory. Some discussions and conclusions are given in section 6.

\section{Lorentz violation theory and the modified Hamilton-Jacobi equation} 
The recent researches on high energy quantum gravity theory show that the Lorentz-violation is needed to be modified in the Planck scale. Gomes, et al. [76] developed a new Lorentz violation scalar equation in which the Lagrangian ${\mathcal{L}}$ is given as
\begin{eqnarray}
{\mathcal{L}}&=&\frac{1}{2}[\partial_\mu\psi\partial^\mu\psi+\lambda(u^\alpha\partial_\alpha\psi)^2+m^2\psi^2],
\end{eqnarray}
where $m$ and $u^\alpha$ are the mass of the scalar particle and etheric like vector respectively and $\lambda$ is a proportionality constant and $u^\alpha$ also satisfies
\begin{eqnarray}
u^\alpha u_\alpha&=&{\rm const.}
\end{eqnarray}
{\bf Coupling of the electromagnetic field with the ether like vector gives rise to a preferred direction in space time leading to Lorentz symmetry violation [77, 78]}. 
The scalar field equation in flat space time in accordance with the principle of least action is
\begin{eqnarray}
\partial_\mu\partial^\mu\psi+\lambda u^\alpha u^\beta\partial_\alpha\partial_\beta\psi+m^2\psi &=&0.
\end{eqnarray}
The etheric-like vector $u^\alpha$ may not be taken as a constant quantity in curved space time but it can hold Eq. (2). The action of scalar particle in which the scalar field theory under Lorentz-violation is taken into consideration can be written as
\begin{eqnarray}
{\mathcal{S}}&=&\int d^4x\sqrt{-g}\frac{1}{2}[\partial_\mu\psi\partial^\mu\psi+\lambda(u^\alpha\partial_\alpha\psi)^2+m^2\psi^2].
\end{eqnarray}
From Eq. (4), the Lorentz-violation of scalar field equation in the curved space time can be written as
\begin{eqnarray}
-\frac{1}{\sqrt{-g}}\partial_\mu[\sqrt{-g}(g^{\mu\nu}+\lambda u^\mu u^\nu)\partial_\nu\psi]+m^2\psi &=&0.
\end{eqnarray}
Eq. (5) is the modified form of Klein-Gordon equation of scalar particle with mass $m$ in the Lorentz-violation theory. In Eq. (5), if the constant term $\lambda$ tends to zero, the original Klein-Gordon equation is recovered. To  derive modified Hamilton-Jacobi equation in the curved space time, the wave function $\psi$ in Eq. (5) can be written as
\begin{equation}
\psi=\psi_c e^{\frac{i}{\hbar}S},
\end{equation}
  where $\mathcal{S}$=$\mathcal{S}$$(t,r,\theta, \phi)$ and $\hbar$ are the Hamiltonian principal function of scalar field and Planck constant respectively. Using Eq. (6) in Eq. (5) and neglecting higher order terms of $\hbar$ in accordance with semiclassical theory, we obtain
\begin{eqnarray}
(g^{\mu\nu}+\lambda u^\mu u^\nu)\partial_\mu S\partial_\nu S+m^2 &=&0.
\end{eqnarray}
Eq. (7) represents not only the dynamical equation of a scalar particle of mass $m$ but also modified form of Hamilton-Jacobi equation of scalar particle in the Lorentz-violation theory of curved space time. This modified Hamilton-Jacobi equation will be used to investigate the entropy, modified Hawking temperature and specific heat capacity of Kerr-de Sitter black hole in the next sections. 
\section{Kerr-de Sitter solution and naive coordinate}
 The Kerr-de Sitter (KdS) black hole is a stationary rotating black hole. The vacuum solution of of KdS black hole in $(t, r, \theta, \phi)$ coordinate system with cosmological constant $\Lambda$ is given by [79]
\begin{eqnarray}
ds^2&=&-\frac{\Delta_r-\Delta_\theta a^2\sin^2\theta}{\rho^2\Xi^2}dt^2+\frac{\rho^2}{\Delta_r}dr^2-\frac{2a[\Delta_\theta(r^2+a^2)-\Delta_r]\sin^2\theta}{\rho^2\Xi^2}dtd\phi
\cr&&
+\frac{\rho^2}{\Delta_\theta}d\theta^2+\frac{\Delta_\theta(r^2+a^2)^2-\Delta_ra^2\sin^2\theta}{\rho^2\Xi^2}\sin^2\theta d\phi^2\,
\end{eqnarray}
where
\begin{eqnarray}
&&\rho^2=r^2+a^2\cos^2\theta,\,\,\,\,\, \Xi=1+\frac{1}{3}\Lambda a^2,\cr
&&\Delta_\theta=1+\frac{1}{3}\Lambda a^2\cos^2\theta,\cr
&&\Delta_r=(r^2+a^2)\Big(1-\frac{1}{3}\Lambda r^2\Big)-2Mr.
\end{eqnarray}
Here  $\Lambda$,  $M$ and $a$ represent the cosmological constant, mass of the black hole and rotating parameter. The event horizon equation of KdS black hole can be obtained from the null hypersurface equation [9, 10 ]
\begin{eqnarray}
g^{\mu \nu}\frac{\partial H}{\partial x^\mu}\frac{\partial H}{\partial x^\nu}&=&0,
\end{eqnarray}
where $H=H(r)$. Then the event horizon equation is 
\begin{equation}
\Delta_r=(r^2+a^2)\Big(1-\frac{1}{3}\Lambda r^2\Big)-2Mr=0.
\end{equation}
Equation (11) has four real roots for $\Lambda>0$ namely $r_+$, $r_h$, $r_-$ and $r_{--}$ $(r_+> r_h> r_->0>r_{--})$. The locations of $r_+$, $r_h$ and $r_-$  represent the cosmological horizon, event horizon and Cauchy horizon respectively. To investigate the Hawking radiation near the event horizon, $\Delta_r$ can be factorised as follows [25]
\begin{eqnarray}
\Delta_r=(r-r_h)\Xi^2L(r), 
\end{eqnarray}
with
\begin{eqnarray}
L(r)=-\frac{\Lambda}{3}\Big[r^3+r_hr^2+\Big(a^2+r^2_h-\frac{3}{\Lambda}\Big)r+\frac{3a^2}{\Lambda r_h}\Big].
\end{eqnarray}
The entropy of KdS black hole near the event horizon is calculated as
\begin{eqnarray}
S_{BH}=\frac{A}{4}=\frac{\pi(r^2_h+a^2)}{\Xi}.
\end{eqnarray}
The KdS black hole has a coordinate singularity near the event horizon. Now the dragging co-ordinate system will be applied in KdS black hole. Let $\frac{d\phi}{dt}=-\frac{g_{14}}{g_{44}}$ [17, 29], the space time metric of KdS black hole (8) reduces to
\begin{eqnarray}
ds^2=\hat{g}_{11}dt^2+\frac{\rho^2}{\Delta_r}dr^2+\frac{\rho^2}{\Delta_\theta}d\theta^2,
\end{eqnarray}
where
\begin{eqnarray}
\hat{g}_{11}=-\frac{\Delta_r\Delta_\theta\rho^2}{\Xi^2[\Delta_\theta(r^2+a^2)^2-\Delta_r a^2\sin^2\theta]}.
\end{eqnarray}
The angular velocity of KdS black hole near the event horizon is defined by
\begin{eqnarray}
\Omega_h=\frac{a}{r^2_h+a^2}.
\end{eqnarray}
Using Eq. (15), the  contravariant components of $g^{ab}$ are derived as

\begin{eqnarray}
\hat{g}^{11}=-\frac{\Xi^2[\Delta_\theta(r^2+a^2)^2-\Delta_r a^2\sin^2\theta]}{\Delta_r\Delta_\theta\rho^2},\,\,\,\,\,\, \hat{g}^{22}=\frac{\Delta_r}{\rho^2},\,\,\,\,\,\hat{g}^{33}=\frac{\Delta_\theta}{\rho^2}.
\end{eqnarray}
The surface gravity near the event horizon of KdS black hole can be obtained as [80, 81]
\begin{eqnarray}
&\kappa=&\lim_{\hat g_{11} \rightarrow 0}\Bigg(-\frac{1}{2}\sqrt{-\frac{g^{22}}{\hat g_{11}}}\frac{d\hat g_{11}}{dr}\Bigg)
=\frac{\Xi(3r_h^2-3\Lambda r_h^4-\Lambda a^2r_h^2-3a^2)}{6r_h(r_h^2+a^2)}.
\end{eqnarray}
The Hawking temperature near the event horizon of KdS black hole  is given by $T_H=\frac{\kappa}{2\pi}$
\begin{eqnarray}
T_H&=&\frac{\Xi(3r_h^2-3\Lambda r_h^4-\Lambda a^2r_h^2-3a^2)}{12\pi r_h(r_h^2+a^2)}.
\end{eqnarray}
To obtain the heat capacity of KdS black hole, the black hole mass can be derived from 
$\Delta_{r}(r_h)=0$ as
\begin{eqnarray}
M&=&\frac{r_h}{2}+\frac{a^2}{2r_h}-\frac{\Lambda r_h^3}{6}-\frac{\Lambda a^2 r_h}{3}.
\end{eqnarray}
The heat capacity $(C_h)$ of blak hole [82, 83] is defined by
\begin{eqnarray}
C_h&=&\frac{\partial M}{\partial T_h}=\Bigg(\frac{\partial M}{\partial r_h}\Bigg)\Bigg(\frac{\partial r_h}{\partial T_h}\Bigg).
\end{eqnarray}
The heat capacity of KdS black hole near the event horizon $r=r_h$ is obtained as
\begin{eqnarray}
C_h&=&\frac{2\pi(r_h^2+a^2)^{2}F_1}{\Xi[F_2(3r_h^2+a^2)-r_h^2(r_h^2+a^2)F_3]},
\end{eqnarray}
where the values of $F_1$, $F_2$ and $F_3$ are defined by
\begin{eqnarray}
F_1&=&3r_h^2-3\Lambda r_h^4-3a^2-2\Lambda a^2r_h^2,\cr
F_2&=&3\Lambda r_h^4+\Lambda a^2r_h^2-3r_h^2+3a^2,\cr
F_3&=&12\Lambda r_h^2+2\Lambda a^2-6.
\end{eqnarray}
The Landau's condition of the coordinate clock synchronization [84] is satisfied by the space-time (15). It is obvious that event horizon and infinite redshift surface of KdS black hole are concordant with each other. This indicates that the geometrical optics limit can be used. The relationship between the imaginary part of the radial action and tunneling probability can be derived using WKB approximation [85]. To study Hawking radiation near the event horizon of KdS black hole, the ether like vector $u^\alpha$ from (15) can be taken as
\begin{eqnarray}
u^t&=&\frac{c_t}{\sqrt{-g_{11}}}=c_t\sqrt{\frac{\Xi^2K}{\Delta_r\Delta_\theta\rho^2}},\cr
u^r&=&\frac{c_r}{\sqrt{g_{rr}}}=\frac{c_r\sqrt{\Delta_{r}}}{\sqrt{\rho^2}},\cr
u^{\theta}&=&\frac{c_{\theta}}{\sqrt{g_{\theta\theta}}}=\frac{c_{\theta}\sqrt{\Delta_\theta}}{\sqrt{\rho^2}},
\end{eqnarray}
where $c_t$, $c_r$, $c_\theta$ are the arbitrary constants and $ K =\Delta_\theta(r^2+a^2)^2-\Delta_ra^2 \sin^2\theta$.  Near the event horizon of black hole, $u^\alpha$ satisfies the following condition as
\begin{eqnarray}
u^\alpha u_\alpha= -c_t^2+c_r^2+c_\theta^2 ={\rm constant}. 
\end{eqnarray}

Again using Eqs. (25) and (26) in Eq. (7), we obtain
\begin{eqnarray}
&&\hat{g}^{11}(\partial_t S)^2+\hat{g}^{22}(\partial_r S)^2+\hat{g}^{33}(\partial_\theta S)^2+\lambda u^tu^t(\partial_t S)^2+\lambda u^ru^r(\partial_r S)^2\cr&&+\lambda u^\theta u^\theta(\partial_\theta S)^2+2\lambda u^tu^r(\partial_t S)(\partial_r S)+2\lambda u^tu^\theta(\partial_t S)(\partial_\theta S)\cr&&+2\lambda u^ru^\theta(\partial_r S)(\partial_\theta S)+m^2=0.
\end{eqnarray}
Eq. (27) contains the three independent variables namely $t$, $r$ and $\theta$. To study the entropy of black hole, the action $S$ can be written as 
\begin{eqnarray}
S=-\omega t+R(r,\theta)+j\phi,
\end{eqnarray}
where $R(r,\theta)$, $\omega$ and $j$ denote the generalized momentum, energy of the emitted particle and  angular momentum with respect to $\phi$-axis respectively.  Using Eq. (27) in  (28), the action $S$ can be written as
\begin{eqnarray}
S&=&-\omega t+\int \frac{-B\pm\sqrt{B^2-4AC}}{2A}dr+j\phi,\cr&
\end{eqnarray}
where 
\begin{eqnarray*}
&A=& \frac{\Delta _r (1+\lambda c_r^2)}{\rho^2},\cr
&B=& 2\lambda\Big[\frac{c_r c_\theta \sqrt{\Delta _r \Delta_\theta}}{\rho^2}\Big(\frac{\partial R}{\partial \theta}\Big)-\frac{c_t c_r \Xi \sqrt{K}(w-j\Omega_h)}{\sqrt{\Delta _\theta }\rho^2}\Big],\cr
&C=&\frac{\Delta_\theta (1+\lambda c_\theta^2)}{\rho^2}\Big(\frac{\partial R}{\partial \theta}\Big)^2-\frac{\Xi^2 K}{\Delta _r \Delta_\theta \rho^2}(\omega-j\Omega_h)^2(1-\lambda c_t^2)\cr&& -2\lambda c_t c_\theta \frac{\Xi \sqrt{K}}{\sqrt{\Delta _r }\rho^2}(\omega-j\Omega_h)\Big(\frac{\partial R}{\partial \theta}\Big)+m^2.
\end{eqnarray*}
The KdS black hole has a pole at the event horizon $r=r_h$. Applying residue theorem of complex analysis and Feynman prescription, the imaginary part of the radial action will be evaluated. The imaginary part of the radial action can be found as
\begin{eqnarray}
{\rm ImS}=\pi\frac{(r^2_h+a^2)(\omega-j\Omega_h)(\lambda c_tc_r+\sqrt{1+\lambda c_r^2-\lambda c_t^2})}{\Xi L(r_h) (1+\lambda c_r^2)}
\end{eqnarray}
Using WKB approximation, the tunneling rate is obtained as
\begin{eqnarray}
\Gamma &\sim&e^{-2{\rm ImS}}.
\end{eqnarray}
Then the Hawking temperature of KdS black hole near the event horizon $r=r_h$ due to naive coordinate in Lorentz violation is given by
\begin{eqnarray}
T_n&=&\frac{\Xi (1+\lambda c_r^2)}{(\lambda c_tc_r+\sqrt{1+\lambda c_r^2-\lambda c_t^2})}\frac{(3r_h^2-3\Lambda r_h^4-\Lambda a^2r_h^2-3a^2)}{6\pi r_h(r_h^2+a^2)}.
\end{eqnarray}
 The tunneling probability and its radiation spectrum can be derived using WKB approximation. Taking the conservation of energy and angular momentum, the mass and angular momentum of black hole is allowed to fluctuate. If the particle of energy  $\omega$ and angular momentum $j$ radiate from the black hole, the  mass and angular momentum of KdS black hole  will be $M-\omega$ and $J-j$ respectively. The imaginary part of the radial action is calculated as follows
\begin{eqnarray}
{\rm ImS} &=&\pi\int^{(\omega, J)}_{(0,0)}\frac{(r^2_h+a^2)}{\Xi L(r_h)}\frac{[\lambda c_tc_r+\sqrt{1+\lambda c_r^2-\lambda c_t^2}]}{ (1+\lambda c_r^2)}(d\omega-\Omega_h dj),\cr
&=&-\pi\frac{[\lambda c_tc_r+\sqrt{1+\lambda c_r^2-\lambda c_t^2}]}{(1+\lambda c_r^2)}\int^{M-\omega}_{M}\frac{r^2_h}{\Xi L(r_h)}dM,
\end{eqnarray}
where $\Omega_h=\frac{a}{r^2_h+a^2}$ and $j=Ma$.  From $\Delta_r=0$, the change of mass is calculated as
\begin{eqnarray}
dM=\Big[-\frac{\Lambda}{2}r^2_h+\Big(\frac{1}{2}-\frac{\Lambda a^2}{6}\Big)-\frac{a^2}{2r^2_h}\Big]dr_h=\frac{L(r_h)}{2r_h}dr_h.
\end{eqnarray}
Then, the Eq. (33) can be written as
\begin{eqnarray}
{\rm ImS} &=&-\frac{\pi(\lambda c_tc_r+\sqrt{1+\lambda c_r^2-\lambda c_t^2}) }{2\Xi(1+\lambda c_r^2)}\int^{r_e}_{r_f}r_hdr_h\cr&&=-\frac{\pi(r^2_e-r^2_f)}{4\Xi}\frac{(\lambda c_tc_r+\sqrt{1+\lambda c_r^2-\lambda c_t^2})}{(1+\lambda c_r^2)},
\end{eqnarray}
where $r_e$ and $r_f$ are the locations of horizons of KdS black hole before and after the particle emission. The tunneling probability can be calculated using WKB approximation as
\begin{eqnarray}
\Gamma &\sim& exp\Bigg[\frac{[\lambda c_tc_r+\sqrt{1+\lambda c_r^2-\lambda c_t^2}]}{(1+\lambda c_r^2)}\Bigg(\frac{\pi(r^2_e-r^2_f)}{2\Xi}\Bigg)\Bigg],\cr&=&exp\Bigg[\frac{[\lambda c_tc_r+\sqrt{1+\lambda c_r^2-\lambda c_t^2}]}{(1+\lambda c_r^2)}\Bigg(\frac{\pi(r^2_e-a^2)}{2\Xi}-\frac{\pi(r^2_f-a^2)}{2\Xi}\Bigg)\Bigg],\cr&=&exp\Bigg[\frac{[\lambda c_tc_r+\sqrt{1+\lambda c_r^2-\lambda c_t^2}]}{(1+\lambda c_r^2)}\Big(\frac{1}{2}\Delta S_{BH}\Big)\Bigg].
\end{eqnarray}
The change in Bekenstein-Hawking entropy of KdS black hole due to Lorentz violation theory in naive coordinate is given by 
\begin{eqnarray}
\frac{1}{2}\frac{[\lambda c_tc_r+\sqrt{1+\lambda c_r^2-\lambda c_t^2}]}{(1+\lambda c_r^2)}\Delta S_{BH}&=&\frac{1}{2}\frac{[\lambda c_tc_r+\sqrt{1+\lambda c_r^2-\lambda c_t^2}]}{(1+\lambda c_r^2)}\cr\times&&[S_{BH}(M-\omega)-S_{BH}(M)].
\end{eqnarray}
 It shows that the change in Bekenstein-Hawking entropy is modified due to presence of Lorentz violation parameter $\lambda$ and ether like vectors $u^\alpha$. When $\lambda=0$, Eq. (37) becomes $\frac{1}{2}\Delta S_{BH}=\frac{1}{2}[S_{BH}(M-\omega)-S_{BH}(M)]$ which is half of correct value of Bekenstein-Hawking entropy near the event horizon of KdS black hole.

 We calculate the heat capacity of KdS black hole at the event horizon $r=r_h$ due to Lorentz violation in naive coordinte as 
\begin{eqnarray}
C_{n}&=&\frac{\pi(\lambda c_tc_r+\sqrt{1+\lambda c_r^2-\lambda c_t^2})(r_h^2+a^2)^{2}F_1}{\Xi(1+\lambda c_r^2)[F_2(3r_h^2+a^2)-r_h^2(r_h^2+a^2)F_3]},
\end{eqnarray}
where the values of $F_1$, $F_2$ and $F_3$ are given in (24). From Eqs. (32), (37) and (38), we observe that the Hawking temperature, entropy and heat capacity depend on Lorentz violation parameter $\lambda$ and ether like vectors $u^\alpha$.
\section{Painleve coordinate}

To discuss the tunneling rate near the event horizon $ r=r_h$ of KdS black hole, the well-defined coordinate system known as Painleve coordinate system is considered. Let
\begin{eqnarray}
dt=dT+F(r,\theta)dr+G(r,\theta)d\theta,
\end{eqnarray}
where $F(r,\theta)$ and $G(r,\theta)$ represent the two arbitrary functions satisfying the following condition
\begin{eqnarray}
\frac{\partial F(r,\theta)}{\partial\theta}=\frac{\partial G(r,\theta)}{\partial r}.
\end{eqnarray}
It is obvious that  the constant-time slice is a flat Euclidean space and taking
\begin{eqnarray}
g_{22}+F^2(r,\theta)\hat{g}_{11}=1,
\end{eqnarray}
 the KdS black hole in Painleve coordinate is found as 
\begin{eqnarray}
ds^2&&=\hat{g}_{11}dT^2+2\sqrt{\hat{g}_{11}\Big(1-\frac{\rho^2}{\Delta_r}\Big)}dTdr+dr^2
+\Big(\frac{\rho^2}{\Delta_\theta}+G^2\hat{g}_{11}\Big)d\theta^2\cr&&+2\hat{g}_{11}GdT d\theta+2G\sqrt{\hat{g}_{11}\Big(1-\frac{\rho^2}{\Delta_r}\Big)}dr d\theta.
\end{eqnarray}
The contravariant components of $g^{ab}$ from Eq. (42) are found as
\begin{eqnarray}
g^{11}&=&\frac{1}{\hat{g}_{11}\rho^2}[\Delta_r+\Delta_\theta G^2\hat{g}_{11}],\,\,\,\,\,\,g^{23}=g^{32}=0,\,\,\,\,\,\,g^{22}=\frac{\Delta_r}{\rho^2},\cr
g^{12}&=&-\frac{\Delta_r}{\hat{g}_{11}\rho^2}\sqrt{\hat{g}_{11}\Big(1-\frac{\rho^2}{\Delta_r}\Big)},\,\,\,\,\,\,g^{31}=-\frac{G\Delta_\theta}{\rho^2},\,\,\,\,\,\,g^{33}=\frac{\Delta_\theta}{\rho^2}.
\end{eqnarray}
To derive the modified Hawking temperature, entropy and  heat capacity near the event horizon of KdS black hole, the ether like vectors $u^{\alpha}$ can be constructed from Eq. (42) as follows
\begin{eqnarray}
u^T&=&\frac{c_{T}}{\sqrt{\hat{g}_{11}}},\cr
u^r&=&\frac{c_{r}\sqrt{\hat{g}_{11}}}{g_{12}}=\frac{c_r\sqrt{\Delta_r}}{\sqrt{(\Delta_r-\rho^2)}},\cr
u^{\theta}&=&\frac{c_{\theta}}{\sqrt{g_{33}}}=\frac{c_{\theta}}{\sqrt{\frac{\rho^{2}}{\Delta_{\theta}}+G^{2}\hat{g}_{11}}},
\end{eqnarray}
where $c_{T}$, $c_r$ and $c_{\theta}$ are arbitrary constants and  $u^{\alpha}u_{\alpha}=c^2_{T}+2 c_Tc_r+c^2_{\theta}={\rm constant}$ (near the event horizon of black hole). Using Eqs. (43) and (44) in Eq. (7), we get
\begin{eqnarray}
&&\frac{1}{\hat{g}_{11}\rho^2}\Big(\Delta_r+\Delta_\theta G^2\hat{g}_{11}\Big)\Big(\frac{\partial S}{\partial T}\Big)^2-\frac{2\Delta_r}{\hat{g}_{11}\rho^2}\sqrt{\hat{g}_{11}\Big(1-\frac{\rho^2}{\Delta_r}\Big)}\Big(\frac{\partial S}{\partial T}\Big)\Big(\frac{\partial S}{\partial r}\Big)\cr&&
+\frac{\Delta_r }{\rho^2}\Big(\frac{\partial S}{\partial r}\Big)^2+\frac{\Delta_\theta }{\rho^2}\Big(\frac{\partial S}{\partial \theta}\Big)^2-\frac{2G\Delta_\theta}{\rho^2}\Big(\frac{\partial S}{\partial \theta}\Big)\Big(\frac{\partial S}{\partial T}\Big)+\lambda u^\mu u^\nu \partial_\mu S\partial_\nu S +m^2=0.\cr&&
\end{eqnarray}

The above equation contains three independent variables namely $T$, $r$ and $\theta$. The Hawking radiation occurs along the radial direction only. To separate the variables from Eq. (45),  we write the action $S$ as follows
\begin{eqnarray}
S=-\omega T+W(r,\theta)+j\phi.
\end{eqnarray}
Using Eqs. (44) and (46) in Eq. (45), a quadratic equation in $(\frac{\partial W}{\partial r})$ is  obtained as follows
\begin{eqnarray}
M\Big(\frac{\partial W}{\partial r}\Big)^2+N\Big(\frac{\partial W}{\partial r}\Big)+P=0,
\end{eqnarray}
where
\begin{eqnarray*}
& M=&\frac{\Delta_r}{\rho^2}+\frac{\lambda c_r^2\Delta_r}{\Delta_r-\rho^2},\cr
 &N=&(w-j\Omega_h)\Bigg(\frac{(2\Delta_r-2\rho^2-2\lambda c_Tc_r\rho^2)\Xi\sqrt{K}}{\rho^2\sqrt{-\Delta_\theta\rho^2(\Delta_r-\rho^2)}}\Bigg)+\cr&&\frac{2\lambda c_rc_\theta\sqrt{\Delta_r\Delta_\theta}}{\sqrt{(\Delta_r-\rho^2)(\rho^2+G^2\hat{g_{11}}\Delta_\theta)}}\Big(\frac{\partial W}{\partial \theta}\Big) ,\cr
 &P=&\frac{\Delta_\theta}{\rho^2}\Bigg(G(w-j\Omega_h)+\frac{\partial W}{\partial \theta}\Bigg)^2+\frac{\Delta_r}{\rho^2}\frac{1}{\hat{g}_{11}}(w-j\Omega_h)^2+\frac{\lambda c_\theta^2\Delta_\theta}{\rho^2+\Delta_\theta G^2\hat{g}_{11}}\Big(\frac{\partial W}{\partial \theta}\Big)^2\cr&&
 +\frac{\lambda c_T^2(w-j\Omega_h)^2}{\hat{g}_{11}}-\frac{2\lambda c_Tc_\theta(w-j\Omega_h)}{\sqrt{\hat{g}_{11}\Big(\frac{\rho^2}{\Delta_\theta}+G^2\hat{g}_{11}}\Big)}\Big(\frac{\partial W}{\partial \theta}\Big)+m^2.
\end{eqnarray*}
Substituting Eq. (47) in Eq. (46), the outgoing action S can be expressed as 
\begin{eqnarray}
S=-\omega T+\int \frac{-N\pm\sqrt{N^2-4MP}}{2M}dr+j\phi.
\end{eqnarray}
On integration of Eq. (48) near the event horizon $r=r_h$ of KdS black hole using residue theorem of complex analysis, the imaginary part of outgoing action S is found as
\begin{eqnarray}
{\rm ImS}=\frac{2\pi(r^2_h+a^2)(\omega-j\Omega_h)(1+\lambda c_T c_r)}{\Xi L(r_h)(1-\lambda c_r^2)}.
\end{eqnarray}
Using WKB approximation, the tunneling rate is 
\begin{eqnarray}
\Gamma &\sim&e^{-2{\rm ImS}}.
\end{eqnarray}
The Hawking temperature is given by
\begin{eqnarray}
&T_p=&\frac{\Xi(1-\lambda c_r^2)(3r_h^2-3\Lambda r_h^4-\Lambda a^2r_h^2-3a^2)}{12\pi[(1+\lambda c_T c_r)] r_h(r_h^2+a^2)}.
\end{eqnarray}

If $\lambda=0$ and $c_T+c_r=0 $, the Lorentz violation has been cancelled and the original Hawking temperature which is similar to the actual one given in Eq. (20) is obtained.

 The imaginary part of the outgoing action  is calculated as
\begin{eqnarray}
{\rm ImS}&=&-2\pi\int^{M-\omega}_{M}\frac{(r^2_h+a^2)(dw-\Omega_h dj)}{\Xi L(r_h)} \frac{(1+\lambda c_T c_r)}{(1-\lambda c_r^2)}.\nonumber\\
\end{eqnarray}
From Eqs. (34) and (52), we have
\begin{eqnarray}
{\rm ImS}=-\frac{\pi(r^2_e-r^2_f)}{2\Xi}\frac{(1+\lambda c_T c_r)}{(1-\lambda c_r^2)}.
\end{eqnarray}
By using Eq. (53) in (50), we get
\begin{eqnarray}
\Gamma &\sim & exp \Bigg[\frac{(1+\lambda c_T c_r)}{(1-\lambda c_r^2)}\Bigg(\frac{\pi(r^2_e-r^2_f)}{\Xi}\Bigg)\Bigg],\cr&=&exp\Bigg[\frac{(1+\lambda c_T c_r)}{(1-\lambda c_r^2)}\Bigg(\frac{\pi(r^2_e-a^2)}{\Xi}-\frac{\pi(r^2_f-a^2)}{\Xi}\Bigg)\Bigg],\cr&=&exp\Bigg[\frac{(1+\lambda c_T c_r)}{(1-\lambda c_r^2)}(\Delta S_{BH})\Bigg].
\end{eqnarray}
The change in Bekenstein-Hawking entropy of KdS black hole due to Lorentz violation theory in Painleve coordinate is given by 
\begin{eqnarray}
\frac{(1+\lambda c_T c_r)}{(1-\lambda c_r^2)}\Delta S_{BH}&=&\frac{(1+\lambda c_T c_r)}{(1-\lambda c_r^2)}\Big[S_{BH}(M-\omega)-S_{BH}(M)\Big]
\end{eqnarray}
  We calculate the heat capacity of KdS black hole near the event horizon $r=r_h$ due to Lorentz violation in Painleve coordinte as
\begin{eqnarray}
C_{p}&=&\frac{ 2\pi(1+\lambda c_T c_r)(r_h^2+a^2)^{2}F_1}{\Xi(1-\lambda c_r^2)[F_2(3r_h^2+a^2)-r_h^2(r_h^2+a^2)F_3]}.
\end{eqnarray}
 The values of $F_1$, $F_2$ and $F_3$ near the event horizon $r=r_h$ of KdS black hole are given in Eq. (24). We observe that the modified Hawking temperature (51), change in Bekenstein-Hawking entropy (55) and heat capacity (56) near the event horizon of KdS black hole depend on the Lorentz violation parameter $\lambda$ and ether like vectors $u^\alpha$. If $\lambda$ tends to zero, then the original Hawking temperature, change in Bekenstein-Hawking entropy and  heat capacity near the event horizon of KdS black hole are recovered.
 \section{ Eddington coordinate}
To find the entropy of KdS black hole near the event horizon $r=r_h$, the well-behaved coordinate system named as Eddington coordinate will be used.  We consider the tansformation
\begin{eqnarray*}
dt=du-\frac{\Xi}{\Delta_r}\sqrt{\frac{[\Delta_\theta(r^2+a^2)^2-\Delta_r a^2\sin^2\theta]}{\Delta_\theta}}dr,
\end{eqnarray*}
then the space time (15) is transformed as
\begin{eqnarray}
ds^2=-\frac{\Delta_r\Delta_\theta \rho^2}{\Xi^2K}du^2+\frac{2\rho^2}{\Xi}\sqrt{\frac{\Delta_\theta}{K}}du dr+\frac{\rho^2}{\Delta_\theta}d\theta^2.
\end{eqnarray}
The contravariant components of $g^{ab}$ of the above equation can be written as
\begin{eqnarray}
g^{11}=0,\,\,g^{12}=g^{21}=\frac{\Xi}{\rho^2}\sqrt{\frac{K}{\Delta_{\theta}}},\,\,g^{22}=\frac{\Delta_r}{\rho^2},\,\,\,g^{33}=\frac{\Delta_{\theta}}{\rho^2}.
\end{eqnarray}
We construct ether like vectors $u^{\alpha}$ from Eq. (57) as 
\begin{eqnarray}
u^u&=&\frac{c_{u}}{\sqrt{g_{11}}}=c_u\sqrt{-\frac{\Xi^2K}{\Delta_r\Delta_\theta\rho^2}},\cr
u^r&=&\frac{c_{r}\sqrt{g_{11}}}{g_{12}}=\frac{c_r}{\rho^2}\sqrt{-\Delta_r\rho^2K},\cr
u^\theta&=&\frac{c_\theta}{\sqrt{g_{33}}}=c_\theta\Bigg( \sqrt{\frac{\Delta_\theta}{\rho^2}} \Bigg),
\end{eqnarray}
where $c_u$, $c_r$ and $c_\theta$ are arbitrary constant and $u^\alpha u_\alpha=c_u^2+2c_r
c_u+c_\theta^2={\rm constant}$. Using Eqs. (58) and (7), we obtain
\begin{eqnarray}
&&\frac{\Delta_r}{\rho^2}\Big(\frac{\partial S}{\partial r}\Big)^2+\frac{2\Xi}{\rho^2}\sqrt{\frac{K}{\Delta_{\theta}}}\Big(\frac{\partial S}{\partial u}\Big)\Big(\frac{\partial S}{\partial r}\Big)+\frac{\Delta_{\theta}}{\rho^2}\Big(\frac{\partial S}{\partial{\theta}}\Big)^2\cr&&+\lambda u^\mu u^\nu \partial_\mu S\partial_\nu S +m^2=0.
\end{eqnarray}
The above equation involves the variables $u$, $r$ and $\theta$. To find radial action of the space time, the action $S$ can be written as
 \begin{eqnarray}
S=-\omega u+R(r)+\Theta(\theta)+j\phi.
\end{eqnarray}
Using Eqs. (59) and (61) in Eq. (60), a quadratic equation in $\Big(\frac{\partial R}{\partial r}\Big)$ is obtained as
\begin{eqnarray}
&&2\lambda c_rc_\theta\frac{\sqrt{-\Delta_r \Delta_\theta}}{\rho^2}\Big(\frac{\partial \Theta}{\partial \theta}\Big)\Big(\frac{\partial R}{\partial r}\Big)-\frac {2\Xi}{\rho^2}\sqrt{\frac{K}{\Delta_\theta}}(\omega-j\Omega_h)(1+\lambda c_uc_r)\Big(\frac{\partial R}{\partial r}\Big)\cr&&+\frac{\Delta_r}{\rho^2}(1-\lambda c_{r}^2)\Big(\frac{\partial R}{\partial r}\Big)^2+\frac{\Delta_\theta}{\rho^2}(1+\lambda c_\theta^2)\Big(\frac{\partial \Theta}{\partial \theta}\Big)^2-\lambda c_u^2\frac{\Xi^2K}{\Delta_r \Delta_\theta \rho^2}(\omega-j\Omega_h)^2\cr&&-2\lambda c_uc_\theta\frac{\Xi}{\rho^2}\sqrt{\frac{-K}{\Delta_r}}(\omega-j\Omega_h)\Big(\frac{\partial \Theta}{\partial \theta}\Big)+m^2=0.
\end{eqnarray}
Solving $\Big(\frac{\partial R}{\partial r}\Big)$ from Eq. (62), the action $S$ can be expressed as  
\begin{eqnarray}
S=-\omega u+\int\frac{-B^{'}\pm\sqrt{B^{'2}-4A^{'} D^{'}}}{2A^{'}} dr+j\phi,\cr&&
\end{eqnarray}
where
\begin{eqnarray*}
A^{'}&=&\frac{\Delta_r}{\rho^2}(1-\lambda c_{r}^2),\cr
B^{'}&=&-\frac {2\Xi}{\rho^2}\sqrt{\frac{K}{\Delta_\theta}}(\omega-j\Omega_h)(1+\lambda c_uc_r)+2\lambda c_rc_\theta\frac{\sqrt{-\Delta_r \Delta_\theta}}{\rho^2}\Big(\frac{\partial \Theta}{\partial \theta}\Big),\cr
D^{'}&=&\Big[\frac{\Delta_\theta}{\rho^2}(1+\lambda c_\theta^2)\Big(\frac{\partial \Theta}{\partial \theta}\Big)-2\lambda c_uc_\theta\frac{\Xi}{\rho^2}\sqrt{\frac{-K}{\Delta_r}}(\omega-j\Omega_h)\Big]\Big(\frac{\partial \Theta}{\partial \theta}\Big)\cr&&-\lambda c_u^2\frac{\Xi^2K}{\Delta_r \Delta_\theta \rho^2}(\omega-j\Omega_h)^2+m^2.
\end{eqnarray*}
Completing the integration of Eq. (63) near the event horizon $r=r_h$ of KdS black hole  using residue theorem of complex analysis, the outgoing action $S$ is found as
\begin{eqnarray}
{\rm ImS}=2\pi\frac{(r^2_h+a^2)(1+\lambda c_uc_r)(\omega-j\Omega_h)}{\Xi L(r_h)(1-\lambda c_r^2)}.
\end{eqnarray}
Using WKB approximation, the tunneling rate is obtained as
\begin{eqnarray}
\Gamma &\sim&e^{-2{\rm ImS}}.
\end{eqnarray}
The Hawking temperature near the event horizon of KdS black hole due to Lorentz violation theory in Eddington coordinate is obtained as
\begin{eqnarray}
T_u&=&\frac{\Xi(1-\lambda c_r^2)(3r_h^2-3\Lambda r_h^4-\Lambda a^2r_h^2-3a^2)}{12\pi r_h(r_h^2+a^2)(1+\lambda c_uc_r)}.
\end{eqnarray}
If $\lambda=0$ and $c_u+c_r=0$,  the Lorentz violation has been cancelled and the original Hawking temperature which is similar to the temperature given in Eq. (20) is obtained.
Using WKB approximation the tunneling rate is obtained as 
\begin{eqnarray}
\Gamma &\sim& exp\Bigg[\frac{(1+\lambda c_u c_r)}{(1-\lambda c_r^2)}\Bigg(\frac{\pi(r^2_e-r^2_f)}{\Xi}\Bigg)\Bigg]\cr&=&exp\Bigg[\frac{(1+\lambda c_u c_r)}{(1-\lambda c_r^2)}\Delta S_{BH}\Bigg].
\end{eqnarray}

The change in Bekenstein-Hawking entropy near the event horizon of KdS black hole due to Lorentz violation theory in Eddington coordinate is given by 
\begin{eqnarray}
&&\frac{(1+\lambda c_u c_r)}{(1-\lambda c_r^2)}\Delta S_{BH}=\frac{(1+\lambda c_u c_r)}{(1-\lambda c_r^2)}\Big[S_{BH}(M-\omega)-S_{BH}(M)\Big].
 \end{eqnarray}
We observe that the modified entropy depends on the Lorentz violation parameter $\lambda$ and ether like vector $u^{\alpha}$. If $\lambda=0$ and $c_u+c_r=0$ in Eq. (68), the Lorentz violation theory has been cancelled. In such case, the actual Bekenstein-Hawking entropy is obtained near the event horizon of KdS black hole (20).

 We calculate the heat capacity  near the event horizon of KdS black hole at the event horizon $r=r_h$ due to Lorentz violation theory in Eddington coordinate as
\begin{eqnarray}
C_{u}&=&\frac{2\pi(r_h^2+a^2)^{2}(1+\lambda c_uc_r)F_1}{\Xi(1-\lambda c_r^2)[F_2(3r_h^2+a^2)-r_h^2(r_h^2+a^2)F_3]}.
\end{eqnarray}
Near the event horizon $r=r_h$ of KdS black hole, the values of $F_1$, $F_2$ and $F_3$  are given in Eq. (24).
The modified heat capacity depends on Lorentz violation parameter $\lambda$ and the ether like vector $u^\alpha$. In the absence of Lorentz violation theory the original heat capacity  is obtained. It is observed that the well-behaved co-ordinate is necesssary to obtain the original Hawking temperature, change in Bekenstein-Hawking entropy and heat capacity of KdS black hole via Hamilton-Jacobi equation.

\section{Discussion and Conclusion}
 The Hawking temperature near the event horizon of KdS black hole derived from the naive coordinate (32), Painleve coordinate (51) and Eddington coordinate (66) can be recast having different conditions of constant terms as follows 
 \begin{eqnarray}
T_c&=&\delta\frac{\Xi(3r_h^2-3\Lambda r_h^4-\Lambda a^2r_h^2-3a^2)}{6\pi r_h(r_h^2+a^2)}.
\end{eqnarray}
Case (i): If
\begin{eqnarray}
 \delta=\frac{ (1+\lambda c_r^2)}{(\lambda c_tc_r+\sqrt{1+\lambda c_r^2-\lambda c_t^2})},
 \end{eqnarray}
 $T_c$ becomes the modified Hawking temperature of KdS black hole given in Eq. (32). If $\lambda=0$ and $c_r=c_t$, $\lambda\neq 0$, the Lorentz-violation is cancelled. In such case, $T_c$ will be two times of the correct Hawking temperature given in Eq. (20). Our result is consistent with earlier literatures [20, 23, 74, 75].

Case (ii): If
\begin{eqnarray}
 \delta=\frac{1}{2}\frac{(1-\lambda c_r^2)}{(1+\lambda c_T c_r)},
  \end{eqnarray}
Eq. (70) is equal to the modified Hawking temperature of KdS black hole given in Eq. (51). If $\lambda=0$ and $c_T+c_r=0$, the Lorentz violation has been cancelled and $T_c$ becomes the original Hawking temperature given in Eq. (20). If $\delta>\frac{1}{2}$, $c_T+c_r<0$, the Hawking temperature near the event horizon of KdS black hole increases $(T_c= T_p >T_H)$ and if $\delta <\frac{1}{2}$, $c_T+c_r>0$, the Hawking temperature decreases $(T_c= T_p <T_H)$, where $T_p$ and $T_H$  denote the modified Hawking temperature (51) and original Hawking temperature (20) respectively near the event horizon of KdS black hole.

Case (iii): If
\begin{eqnarray}
 \delta=\frac{1}{2}\frac{(1-\lambda c_r^2)}{(1+\lambda c_uc_r)},
 \end{eqnarray}
$T_c$ is the modified Hawking temperature of KdS black hole given in Eq. (66). If $\lambda=0$ and $c_u+c_r=0$,  the Lorentz violation has been cancelled and $T_c$ tends to the original Hawking temperature of KdS black hole given in Eq. (20). If $ \delta>\frac{1}{2}$, $c_u+c_r<0$, the Hawking temperature near the event horizon of KdS black hole increases $(T_c= T_u>T_H)$ and if $ c_u+c_r>0$, the Hawking temperature decreases $(T_c= T_u<T_H)$, where $T_u$ denotes the modified Hawking temperature near the event horizon of KdS black hole given by Eq. (66). The modified Hawking temperatures in the three different coordinate systems and the original Hawking temperature of KdS black hole are plotted with the radius of event horizon in Fig. 1.

\begin{figure}
\centering
\includegraphics{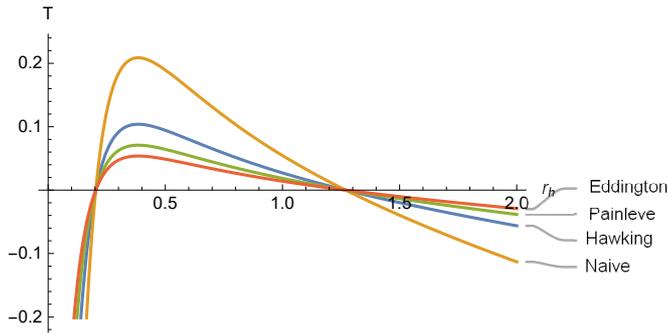}
\caption{Plot of modified Hawking temperature, $T$ with radius of event horizon, ${r_h}$ of KdS black hole for different coordinate systems- naive, Painleve and Eddington. The original Hawking temperature is also shown. Here, $a=0.2$, $\Lambda=0.6$, $\lambda=1$, $c_t=0.4$, $c_r=0.5$, $c_T=0.2$, $c_u=0.9.$}
\end{figure}

 {\bf To investigate the Hawking temperature of Schwarzschild  black hole in static coordinate using the Hamilton-Jacobi equation, the temperature is obtained as $T=1/(4 \pi M)$ which is different from actually calculated by other methods [21, 86, 87]. It is called factor of 2 problem [88]. If Hamilton-Jacobi equation is used to investigate the Hawking temperature of black hole in Painleve coordinate, the correct Hawking temperature is obtained [89]. It shows that the factor of 2 problem is a problem depending upon the choice of coordinate. Akhmedov et al. [90] showed that the factor of 2 problem can not be solved because it may be seen from the fact that the value of the integral should not be changed by simple changing spatial variable.  The same observations can be seen in the study of Hawking radiation and change in  Bekenstein-Hawking entropy of KdS black hole using Hamilton-Jacobi equation in Lorentz violation theory}.

 The change in Bekenstein-Hawking entropies near the event horizon of KdS black hole derived from naive coordinate (37), Painleve coordinate (55) and Eddington coordinate (68) can be combined into a single expression as follows
 \begin{eqnarray}
\beta\Delta S_{BH}=\beta\Big[S_{BH}(M-\omega)-S_{BH}(M)\Big]
 \end{eqnarray}
Case (i): If
 \begin{eqnarray}
 \beta=\frac{1}{2}\frac{[\lambda c_tc_r+\sqrt{1+\lambda c_r^2-\lambda c_t^2}]}{(1+\lambda c_r^2)},
 \end{eqnarray}
Eq. (74) is the change in Bekenstein-Hawking entropy given in Eq. (37). If $\lambda =0$ and $c_r=c_t$, $\lambda\neq 0$, then Eq. (74) becomes  $\frac{1}{2}\Delta S_{BH} =\frac{1}{2}S_{BH}(M-w)-\frac{1}{2}S_{BH}(M),$ which gives half of the Bekenstein-Hawking entropy near the event horizon of KdS black hole [25].

Case (ii): If
 \begin{eqnarray}
 \beta= \frac{(1+\lambda c_T c_r)}{(1-\lambda c_r^2)},
 \end{eqnarray}
 Eq. (74) becomes the change in Bekenstein-Hawking entropy given in Eq. (55). When $\lambda=0$ and  $c_T+c_r=0$, the Lorentz violation has been cancelled  and  Eq. (74) becomes $\Delta S_{BH} =S_{BH}(M-w)-S_{BH}(M),$ which is the actual value of change in Bekenstein-Hawking entropy of KdS black hole. The Bekenstein-Hawking entropy (BHE) of KdS black hole increases if $c_T+c_r>0$ $(\beta>1)$ and the BHE decreases if $c_T+c_r<0$ $(\beta<1)$. \nolinebreak 
 
Case (iii): If
 \begin{eqnarray}
 \beta= \frac{(1+\lambda c_u c_r)}{(1-\lambda c_r^2)},
 \end{eqnarray}
 Eq. (74) is reduced to the change in Bekenstein-Hawking entropy  given in Eq. (68).
When $\lambda=0$ and $c_u+c_r=0$, the Lorentz violation has been cancelled. Then Eq. (74) becomes $\Delta S_{BH} =S_{BH}(M-w)-S_{BH}(M),$ which gives the actual change in Bekenstein-Hawking entropy of KdS black hole. The Bekenstein-Hawking entropy (BHE) of KdS black hole increases if $c_u+c_r>0$ $(\beta>1)$  and the BHE decreases if $c_u+c_r<0$  $(\beta<1)$.

 The heat capacity near the event horizon of KdS black hole derived from naive coordinate (38), Painleve coordinate (56) and Eddington coordinate (69) can be written into a single expression as follows
 \begin{eqnarray}
C_{c}&=&\gamma\frac{\pi(r_h^2+a^2)^{2}F_1}{\Xi[F_2(3r_h^2+a^2)-r_h^2(r_h^2+a^2)F_3]}.
\end{eqnarray}

Case (i): If 
\begin{eqnarray}
\gamma=\frac{(\lambda c_tc_r+\sqrt{1+\lambda c_r^2-\lambda c_t^2})}{(1+\lambda c_r^2)},
\end{eqnarray}
Eq. (78) is the modified heat capacity of KdS black hole given in Eq. (38). If $\lambda=0$ and $c_r=c_t$, $\lambda\neq 0$, the Lorentz-violation theory is cancelled. In such case, the value of heat capacity, $C_c$ is the half of the correct one given in Eq. (23).\nolinebreak
 
Case (ii): If
\begin{eqnarray}
\gamma=\frac{2(1+\lambda c_T c_r)}{(1-\lambda c_r^2)},
\end{eqnarray}
 Eq. (78) represents the  heat capacity of KdS black hole given in Eq. (56). If $\lambda=0$ and $c_T+c_r=0$,  the Lorentz violation has been cancelled and Eq. (78) is equal to the heat capacity of KdS black hole given in Eq. (23). The  heat capacity near the event horizon of KdS black hole  increases, $C_c=C_p>C_h$ if  $c_T+c_r>0$ $( \gamma>2)$   and the heat capacity decreases, $C_c=C_p<C_h$ if $c_T+c_r<0$ $( \gamma<2)$, where $C_p$ and $C_h$ are the modified  heat capacity (56) and original specific heat (23) respectively near the event horizon of KdS black hole.
 
Case (iii): If
\begin{eqnarray}
\gamma= \frac{2(1+\lambda c_uc_r)}{(1-\lambda c_r^2)},
\end{eqnarray}
 Eq. (78) becomes the heat capacity of KdS black hole given in Eq. (69). If $\lambda=0$ and  $c_u+c_r=0$,  the Lorentz violation has been cancelled and $C_c$ tends to the original  heat capacity given in Eq. (23). The  heat capacity near the event horizon of KdS black hole increases, $C_c=C_u>C_h$ if $c_u+c_r>0$ $( \gamma>2)$   and the heat capacity decreases, $C_c=C_u<C_h$ if $c_u+c_r<0$ $( \gamma<2)$, where $C_u$ represents the modified  heat capacity (69) near the event horizon of KdS black hole.

In this paper, we discuss the tunneling of scalar particles across the event horizon of KdS black hole using Klein-Gordon equation with the Lorentz violation theory of curved space time in three co-ordinate systems namely naive coordinate system, Painleve coordinate system and Eddington coordinate system.

The Hawking temperatures, entropies and heat capacities are modified due to the presence of Lorentz violation parameter $\lambda$ and ether like vectors $u^\alpha$. It is shown that the Hawking temperatures, entropies and  heat capacities increase or decrease depending upon the choices of ether like vectors $u^\alpha$. When Lorentz violation theory is cancelled, the naive coordinate gives the incorrect result of Hawking temperature, Bekenstein-Hawking entropy and specific heat capacity but Painleve coordinate and Eddington coordinate give the exact Hawking temperatures, entropies and  heat capacities of KdS black hole.

The variation between naive and Painleve or Eddington coordinate may be seen in a curved manifold, a covariant distribution of the form $1/(r+i0)$ cannot be derived from the non-locally integrable function $1/r$.

 In the Painleve coordinate and Eddington coordinate systems, the metric components are analytic at the event horizon $r=r_h$ of the black hole and there is a time-like Killing vector, which keeps the space time stationary.

Hence the well-behaved coordinates are suitable to study the Hawking radiation, Bekenstein-Hawking entropy and  heat capacity of KdS black hole near the event horizon via Klein-Gordon equation.

\section*{Acknowledgements} 
This work is supported by the DST INSPIRE fellowship, New Delhi, India (Grant No. IF190759). The authors would like to thank the anonymous reviewers for valuable comments and suggestions to improve the paper.

\section*{Competing Interests} 
The authors declare that there is no conflict of interests
regarding the publication of this paper
\section*{Data Availability}
No data were used to support the
findings of this study.

\end{document}